\documentstyle[aps,prl,multicol,epsfig]{revtex}
\begin{document}
\draft
\title{Exact solution, scaling behaviour
and quantum dynamics of a model of an atom-molecule Bose-Einstein
condensate} 

\author { 
Huan-Qiang Zhou\cite{email0}, Jon Links and  Ross
H. McKenzie}

\address{Centre for Mathematical Physics, The University of Queensland,
		     4072, Australia }

\maketitle

\vspace{10pt}

\begin{abstract}
We study the exact solution for a two-mode model describing coherent
coupling between atomic and molecular Bose-Einstein condensates (BEC), in the
context of the Bethe ansatz. By combining an asymptotic and numerical
analysis, we identify the scaling behaviour of the model and determine
the zero temperature expectation value for the coherence and average atomic
occupation. The threshold coupling for production of the molecular BEC is
identified as the point at which the energy gap is minimum. 
Our numerical results indicate a parity effect for the
energy gap between ground and first excited state depending on whether
the total atomic number is odd or even. The numerical calculations for
the quantum dynamics reveals a smooth transition from the atomic to the
molecular BEC.
\end{abstract}

\pacs{PACS numbers: 03.75.Fi, 05.30.Jp}


\def\aa{\alpha} 
\def\bb{\beta}
\def\a{\hat a}
\def\b{\hat b}
\def\d{\dagger}
\def\de{\delta} 
\def\e{\epsilon}
\def\g{\gamma}
\def\K{\kappa}
\def\ap{\approx}
\def\l{\lambda}
\def\o{\omega}
\def\t{\tilde{\tau}}
\def\s{S}
\def\D{\Delta}
\def\L{\Lambda}
\def\T{{\cal T}}
\def\TT{{\tilde{\cal T}}}
\def\E{{\cal E}} 

\def\beq{\begin{equation}}
\def\eeq{\end{equation}}
\def\bea{\begin{eqnarray}}
\def\eea{\end{eqnarray}}
\def\ba{\begin{array}}
\def\ea{\end{array}}
\def\no{\nonumber}
\def\le{\langle}
\def\re{\rangle}
\def\lt{\left}
\def\rt{\right}
\def\o{\omega}
\def\d{\dagger}
\def\nn{\nonumber}
\def\j{{ {\cal J}}}
\def\n{{\hat n}}
\def\N{{\hat N}}
\def\T{{\cal T}}
\def\TT{{\tilde {\cal T}}}

\newcommand{\reff}[1]{eq.~(\ref{#1})}

\begin{multicols}{2}

After the experimental realization of a Bose-Einstein condensate
(BEC) in
dilute alkali gases \cite{and}, many physicists started to 
consider the possibility of producing a molecular Bose-Einstein
condensate from
photoassociation and/or the Feshbach resonance
of an atomic Bose-Einstein condensate
\cite{wynar,inouye}. As discussed by Zoller \cite{zoller}, this
tantalizing problem is now coming toward resolution. 
Donley {\it et al.} \cite{wieman} recently reported the creation of a BEC 
of coherent superpositions of atomic and molecular $^{85}$Rb states. 
This achievement is significant in that the entangled  
state is comprised of two chemically distinct components.   

In anticipation of this result, this novel area
has attracted  considerable attention from theoretical
physicists \cite{drum,java,timm,hein,abee,will,vardi,hope,holland}.
Drummond {\it et al.} \cite{drum,hein} emphasized the finite-dimensionality
of the system and the importance of quantum fluctuation.
Javanainen {\it et al.} \cite{java} systematically analysed the efficiency of
photoassociation of an atomic condensate into its molecular counterpart.
In Refs. \cite{java,timm,hein}, large-amplitude
coherent
oscillations between an atomic BEC and a molecular BEC were predicted
through the use of the Gross-Pitaevski (GP) mean-field theory (MFT).
Others have gone beyond the GP-MFT \cite{hope,holland}. 
Vardi {\it et al.} \cite{vardi} suggested that the large-amplitude atom-molecule
coherent oscillations are damped by the rapid growth
of fluctuations near the dynamically unstable molecular mode, which
contradicts
the MFT predictions. This has caused some disagreement regarding the
atom-molecule conversion and the nature of coherence 
\cite{wieman}. In order to clarify the 
controversies raised by these investigations, it is highly desirable to
extract
some rigorous and analytical results. 

The aim of this Letter is to show that a model Hamiltonian
to describe coherent coupling
between atomic and molecular BEC's described by single modes
is exactly solvable in the
context of
the algebraic Bethe ansatz \cite{kib}. This makes it 
feasible to
apply techniques well-established in the mathematical physics 
literature to
study the
physics behind this new  phenomenon. 
The Bethe ansatz equations are analysed and solved asymptotically in the 
limits of the stable molecular regime and the stable 
atomic regime. Numerical solutions are also obtained for the crossover
regime. We identify a scaling invariance for the model and  
show that the exact solution predicts a smooth transition from a
quasi-periodic and stable regime to a coherent, large-amplitude,
non-periodic oscillating regime. For finite particle number 
there are no stationary points in contrast to the  prediction by
Vardi {\it et al.} using a linearization scheme \cite{vardi}. Instead, the
onset of strong entanglement as the detuning is decreased is identified as the
point at which the energy gap to the first excited state 
takes the minimum value. 

The Hamiltonian takes the form
\beq
H= \frac {\delta}{2} {\hat a}^\dagger {\hat a} +\frac {\Omega}{2}
 ({\hat a}^\dagger {\hat a}^\dagger {\hat b}
  +{\hat b}^\dagger {\hat a}{\hat a}),
  \label{ham}
  \eeq
  where ${\hat a}^\dagger$ and ${\hat b}^\dagger$ denote the creation
  operators
  for atomic and molecular modes, respectively, $\Omega$ is a
  measure of the strength of the matrix element for creation and
  destruction of molecules, and $\delta$ is the molecular binding energy
  in the absence of coupling. A similar Hamiltonian was first used
  to describe optical second harmonic generation \cite{walls} and
  (with additional damping terms) photon squeezing experiments in a
  two-mode interferometer \cite{chat}.
  It was recently investigated by Vardi {\it et al.} \cite{vardi} as
  a model for atom-molecule BEC's. Note that the total atom number operator
$\N= {\hat n}_a+2{\hat n}_b$, where 
${\hat n}_a={\hat a}^\dagger {\hat a},\,
{\hat n}_b=  {\hat b}^\dagger {\hat b}$, provides
  a good quantum number since $[H,\,{\hat N}]=0.$

{\it Bethe ansatz solution.} 
Following the procedure 
of the algebraic Bethe ansatz 
\cite{kib}, we have derived the Bethe ansatz 
equations (BAE) which solve the model (\ref{ham}). 
For given quantum number $M$ the BAE for the spectral parameters
$\{ v_i \}$ read 
\beq
\frac{\delta}{\Omega} - v_i + \frac {2k}{v_i} 
= 2 \sum ^M_{j \neq i} \frac {1}{v_j
-v_i}. \label {bae}
\eeq
Above, $k$ is a discrete
variable which takes values $1/4$ or $3/4$, dependent on whether the total
number of particles is even or odd respectively. More specifically we
have $N=2M+2(k-1/4)$. For each $M$ there are $2M+1$ families of solutions
$\{ v_i \}$ 
to (\ref{bae}). 
For a given solution,   
the corresponding energy eigenvalue is 
\bea E&=&\delta M+\delta(k-1/4)-\Omega\sum_{i=1}^M v_i.  
\label{nrg} \eea 
Although we will not derive the BAE here, we remark that the model is
obtained through a product of the Lax operators for the $su(1,1)$ algebra
\cite{jurco,rktb} and the Heisenberg algebra \cite{kt89} in the quasi-classical limit.
The construction is similar to that used in the solution of the generalized
Tavis-Cummings model from quantum optics \cite{jurco,rktb}. The eigenstates are
also obtained through this procedure. Consider the following
class of states 
\beq \left|v_1,...,v_M\right>=\prod_{i=1}^MC(v_i)\left| 
\Psi\right> \label{estates} \eeq  
where $C(v)=(v\b^\d-\a^\d\a^\d/2)$, 
$\left|\Psi\right>=\left|0\right>$ for $k=1/4$ and
$\left|\Psi\right>=\a^\d\left|0\right>$ for $k=3/4$. 
In the case when the set of parameters $\{v_i\}$ satisfy the BAE then
(\ref{estates}) 
are precisely the eigenstates of the Hamiltonian. 

{\it Asymptotic analysis.} In the limit of large $|\de/\Omega|$ we can perform
an asymptotic analysis of the Bethe ansatz equations to determine the
asymptotic form of the energy spectrum.  
We choose the following ansatz for the Bethe roots
\bea v_i&\ap& \mu_i\Omega/\delta  
~~~~~~~~~~~~~~~~~~~~~~~~~~~~~~i\leq m, \nn \\
v_i&\ap&\delta/\Omega +\e_i+\mu_i\Omega/\delta ~~~~~~~~~~~~~~~i>m. \nn \eea
Substituting into the BAE and solving for $\e_i,\,\mu_i$ leads to the
following asymptotic result for the energy spectrum (cf. \cite{zlmx})   
\bea
E_m&\ap&
\delta (m+k-1/4) \nn \\  
&&~~~~~~+
\Omega^2(3m^2-m+4km-2kM-2mM)/\delta.   \nn
\eea
This result can be confirmed by second-order perturbation theory. 
The level spacings 
$\Delta_m=E_m-E_{m-1}$ are found to be 
\bea \Delta_m&\ap& \delta-2\Omega^2(M+2-3m-2k)/\delta \nn
. \eea

We let $\E$ denote the ground state energy and $\Delta$ the
gap to the first excited state. Employing the Hellmann-Feynman theorem
we can determine the asymptotic form of the following zero temperature
correlations 
$$\left<{\hat n}_a\right>=2\frac{\partial\E}{\partial \delta}, ~~~~
\theta=2\frac{\partial \E}{\partial \Omega}  $$   
where $\theta \equiv -\left<{\hat a}^\dagger {\hat a}^\dagger {\hat b}
  +{\hat b}^\dagger {\hat a}{\hat a}\right>$ 
  is the {\it coherence correlator}. 
We then have for large $\delta/(\Omega N^{1/2})> 0$   
$$ 
\frac{\Delta}{\Omega N^{1/2}}\ap\frac{\delta}{\Omega N^{1/2}}
-\frac{\Omega N^{1/2}}{\delta},~~~     
\frac{\left<n_a\right>}{N}\ap 0, ~~~    
\frac{\theta}{N^{3/2}}\ap 0   $$       
while for large $\delta/(\Omega N^{1/2})<0$ 
$$ 
\frac{\Delta}{\Omega N^{1/2}}\ap -\frac{\delta}{\Omega N^{1/2}}
-2\frac{\Omega N^{1/2}}{\delta}, ~~    
\frac{\left<n_a\right>}{N}\ap 1-\frac{\Omega^2 N}{2\delta^2},$$
$$     
\frac{\theta}{N^{3/2}}\ap -\frac{\Omega N^{1/2}}{\delta}.  $$ 
The above results suggest that the model has scale invariance with respect
to the single variable $\Omega N^{1/2}/\delta$. They also 
suggest that the scaled gap $\Delta/(\Omega N^{1/2})$ will have a minimum
at some positive value of  $\Omega N^{1/2}/\delta$  of the order of unity. 
However, as the
above is an asymptotic result, we need to undertake a numerical analysis
to obtain a more precise picture for the region of small 
$|\delta/(\Omega N^{1/2})|$,
and to establish that the scaling is also valid
in this region.

{\it Numerical analysis.} 
There is a convenient method to determine the energy spectrum without
solving the BAE. This is achieved by resorting to the functional Bethe
ansatz \cite{wiegmann}. Let us introduce the polynomial function
whose zeros are the roots of the BAE; viz.
$$G(u)=\prod_{i=1}^M(1-u/v_i). $$
It can be shown from the BAE that $G$
satisfies the differential equation
$$uG''-(u^2-\delta u-2k)G'+(Mu-E+\delta(k-1/4))G=0$$
subject to  the inital conditions
$G(0)=1,~~~~G'(0)={E-\delta(k-1/4)}/{2k}.$
By setting $G(u)=\sum^M_{n=0} g_n u^n $ the recurrence relation
$$ g_{n+1}=\frac{E-\de (n+k-1/4)}{\Omega (n+1)(n+2k)} g_n
+\frac{n-M-1}{(n+1)(n+2k)}
g_{n-1} $$ 
is readily obtained. It is clear from this relation that $g_n$ is a
polynomial in $E$ of
degree $n$. We also know that $G$ is a polynomial function of degree
$M$ and so we must have $g_{M+1}=0$. The $M+1$ roots of $g_{M+1}$
are precisely the energy levels $E_m$.
Moreover, the eigenstates (\ref{estates}) are expressible as (up to
overall normalisation)
\beq \left|v_1,...,v_M\right>=\sum_{n=0}^Mg_n(\b^\d)^{(M-n)}(\a^\d\a^\d/2)^n
\left|\Psi\right>. \label{gexp} \eeq 

We have implemented these results in order to numerically solve the energy
spectrum for various ranges of the coupling parameters. The numerical
results show that good  scaling behaviour holds for the entire range  of the
couplings, even for small particle numbers. 
A typical example for the coherence
correlator is shown below. 
We remark that the result for the atomic occupation 
shown in the inset is in qualitative agreement with
Fig. 1 of \cite{kmcj}.

\begin{figure}[h]
\centerline{
\psfig{file=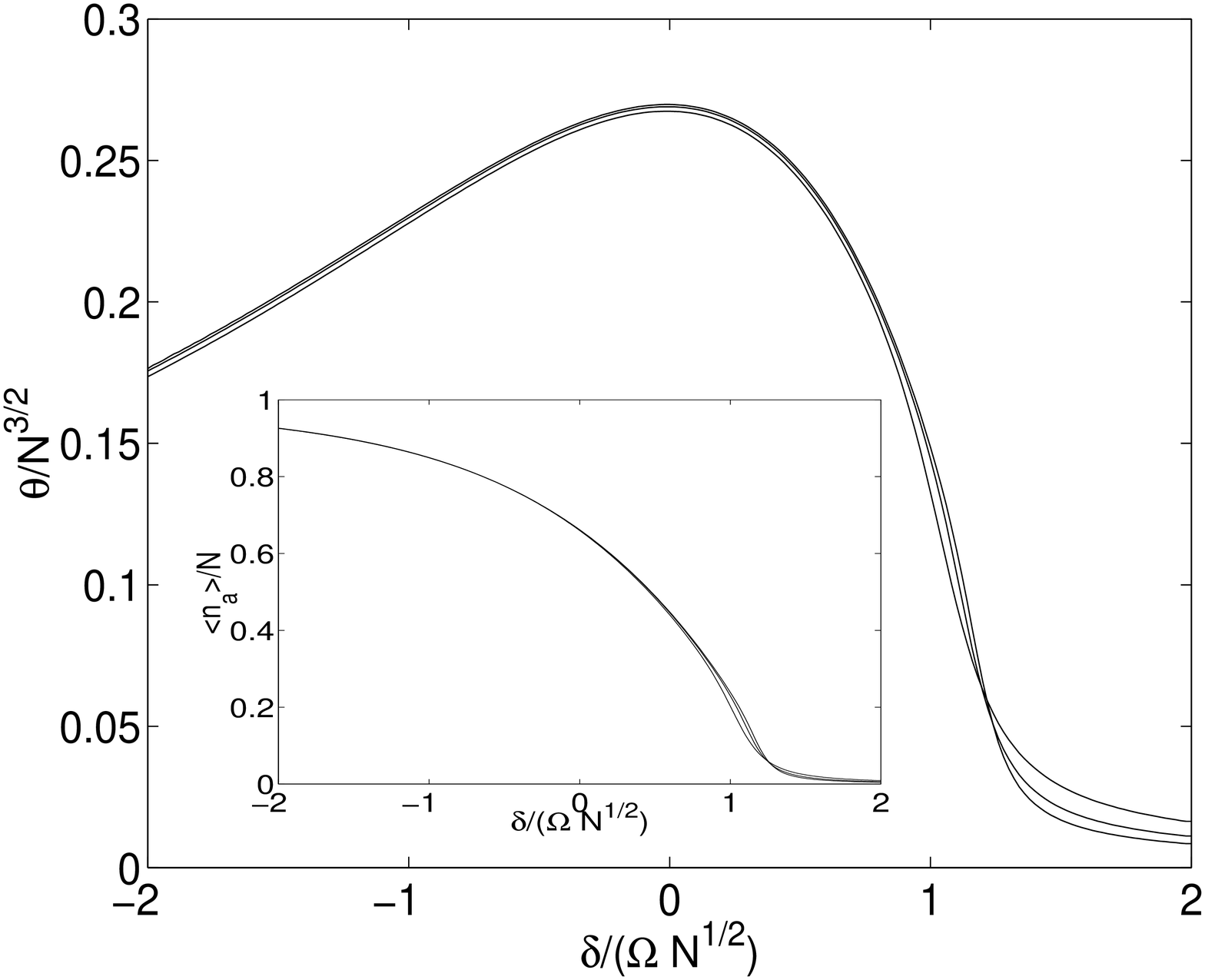,width=7.5cm}}
\caption{
Scaling behaviour of the coherence correlator for 
the ground state as a function of the scaled detuning  $\delta$.
The curves shown are for a total atom number
of $N=20,~30$ and 40.
The threshold coupling for formation of the
predominantly molecular BEC state is $\delta/(\Omega {\sqrt N}) \ap 1.4$,  
indicating
that there is a wide range of the scaled detuning $\delta$
below threshold for which the ground state consists
of a coherent superposition of the atomic and
molecular states. The inset shows the average fractional occupation 
of the atomic state in the ground state.
}
\label{fig1}
\end{figure}

Some striking features which we have
observed are that for fixed $N$ 
there are no level crossings in the energy spectrum
over the entire range of couplings, and 
the existence of a parity effect for the size of the gap 
dependent on whether the
total atomic number $N$ is odd or even, as 
illustrated in Fig. 2.  

The analysis of Vardi {\it et al.} \cite{vardi} in the limit of large $N$ 
has shown, for an entire population of molecular modes, 
the existence of two
stationary  points $|\delta/\Omega|=\sqrt{2N}$,    
and for $|\delta/\Omega|<\sqrt{2N}$
this state becomes unstable.  
Our numerical analysis for finite $N$ shows that a similar situation
occurs but the gap never vanishes.  
From Fig. 2 we see that the minimum value for
$\Delta/\delta$ occurs very close to $\delta=\Omega\sqrt{2N}$.   
 For $\delta/\Omega> \sqrt{2N}$ the state consisting entirely of molecules is
approximately the ground state which is stable due to the large gap to
the next excited level. (The same argument also applies to the negative
$\delta/\Omega$, which simply follows from the inherent symmetry in the
Hamiltonian (\ref{ham})).

\begin{figure}[h]
\centerline{
\psfig{file=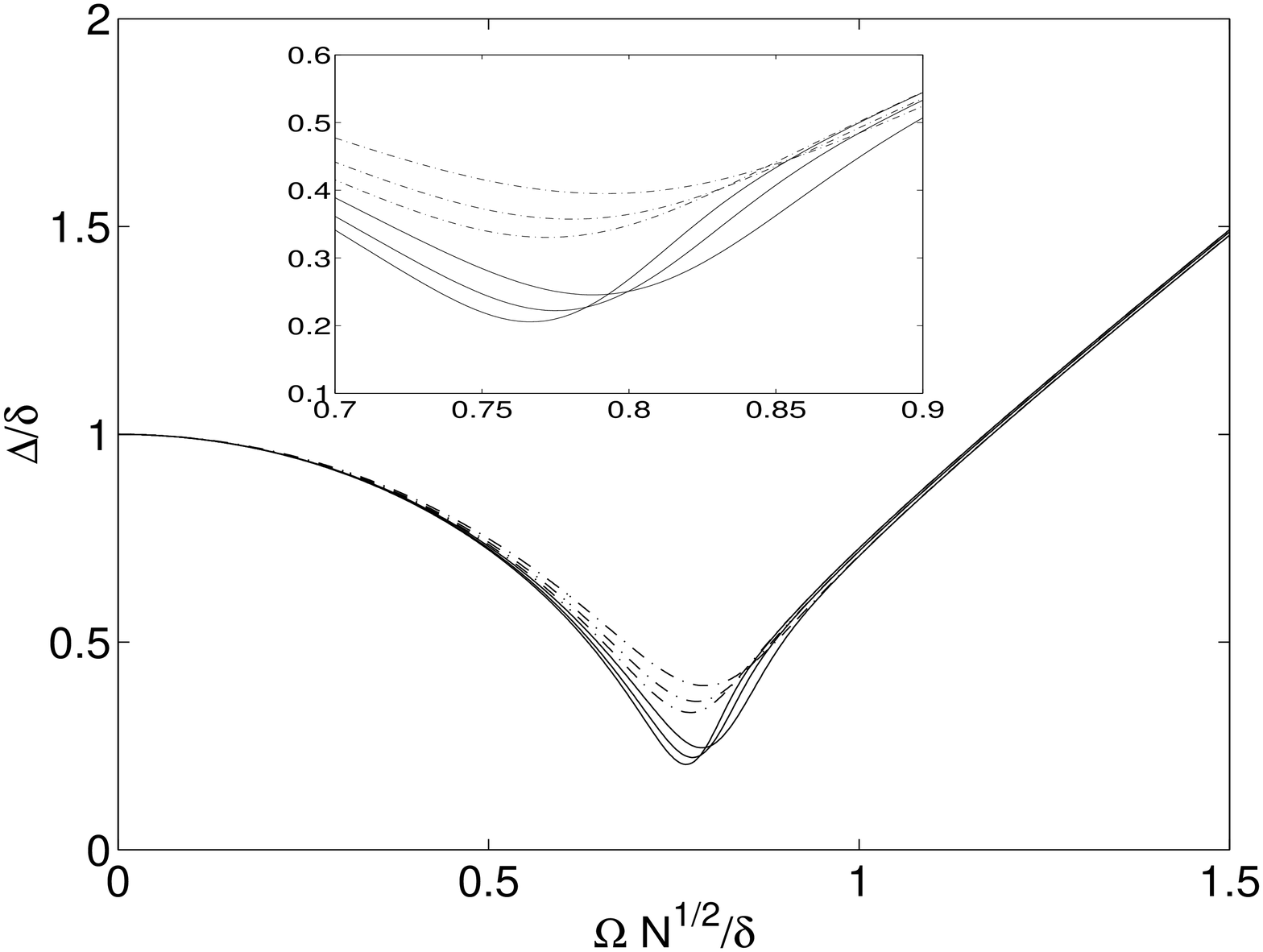,width=7.5cm}}
\caption {
Reduction in the energy gap to the first excited
state near the threshold coupling for formation of
the predominantly molecular BEC. In other words, near threshold
quantum fluctuations lead to a significant reduction
in the binding energy of the molecules.
However, note that the energy gap never vanishes for finite $N$,
in contrast to the treatment of Vardi {\it et al.}. 
\protect\cite{vardi}.
The inset shows a significant parity effect.
Near threshold the gap is larger for an odd number
of atoms than for an even number of atoms.
The solid (dash) curves shown are for $N=70$, 90, 110
(71, 91, 111).}
\label{fig2}
\end{figure}

\def\l{\left}
\def\r{\right}
{\it Quantum dynamics.} For an initial state $\l |\phi( 0)\r >$, the time
evolution is given by $\l|\phi(t)\r > = U(t)\l |\phi(0)\r>$, with 
$U(t) = \sum^M_{m=0} \left|m\r>\l<m\r| \exp (-i E_mt)$ where $\l|m\r>$ is
the eigenstate with energy $E_m$. We also let $g_n(m)$ denote the
co-efficients in (\ref{gexp}) for $\l|m\r>$. Setting
${\hat s} \equiv (2 \b^\dagger b- \a^\dagger \a)/N$, and taking the 
initial state as the  
 pure molecular state; i.e, $\l|\phi(0)\r> =(1/{\sqrt M!}) (\b^\dagger)^M
\l|\psi\r>$, then we have
\bea 
\l<{\hat s} (t)\r> &=& 1 - (4k-1)/N \nn \\
&&-\frac{8}{N}\sum _{m \neq  m'} {\tilde g}_0 (m) {\tilde g}_0 (m')
\sin ^2 (\Delta_{mm'}t/2) \nn \\
&&\times \sum _{n=0}^M 
(n+k) {\tilde g}_n (m) {\tilde g}_n (m'),
\eea 
with
$$
{\tilde g}_n =\frac { 2^{-n} g_n \sqrt {(M-n)!(2n+2k-\frac {1}{2})! }}
{\sqrt { \sum _{m=0}^M 2^{-2m} (M-m)!(2m+2k -\frac {1}{2})! g^2_m}}
$$
and $\Delta_{mm'}=E_m-E_{m'}$.   
Similarly, if the initial state is a pure atomic state, then
\bea 
\l <{\hat s}(t)\r > &=& -1 - \frac{8}{N}  
\sum _{m \neq  m'} {\tilde g}_M (m) {\tilde g}_M (m')
\sin ^2 (\Delta_{mm'}t/2) \nn \\
&&\times \sum _{n=0}^M 
(M-n) {\tilde g}_n (m) {\tilde g}_n (m').
\eea

From the expression for $\l<{\hat s}(t)\r>$, we see that the short time behavior is
quadratic rather than linear in $t$, due to the square of the sine
functions, which is qualitatively
consistent with the results of Vardi {\it et al.} \cite{vardi} using a
linearized model. As for the long time behavior, one can see that
(i) when $N=2$ and 3, there are only two levels, so it is always periodic.
(ii) when $\delta/\Omega$ is large and positive, 
the ground state and first excited state dominate the dynamics,  
since in this regime, the molecular state is approximately the ground state. 
As the energy levels are almost equidistant, the evolution is  
quasi-periodic (with period $T\approx 2\pi/\Delta$) and stable. A similar
situation prevails for largely negative $\delta/\Omega$. 
(iii) when $\delta/\Omega$ is small, all the eigenstates are involved in the 
evolution due to strong
entanglement. This means, all
$g_n (m)$  are of the same order
for all $\l|m\r>$. In this regime, the levels are not
uniformly distributed, so $\l<{\hat s}(t)\r>$ is  a non-trivial sum of  functions
$\sin^2(\Delta_{mm'}t)$ 
with coefficients depending on $g_n(m)$ and $g_n(m')$.  Therefore, the
evolution is not
quasi-periodic. 

\begin{figure}[h]
\centerline{
\psfig{file=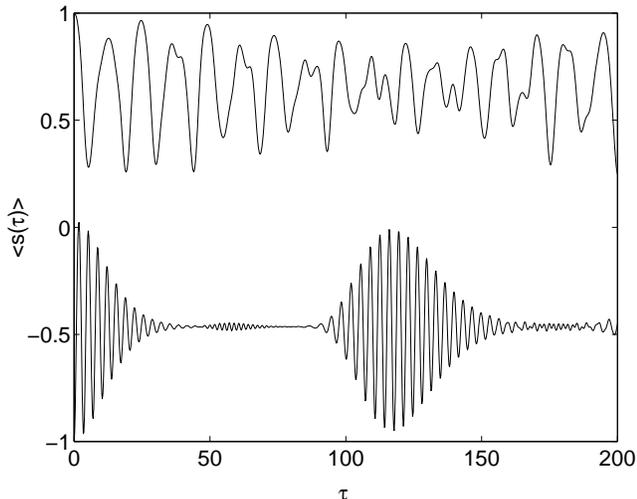,width=8.5cm}}
\caption{
Time evolution of the average relative atom number
$\left<{\hat s}(\tau)\right>$,
where $\tau=\Omega \sqrt{N}t$ is the scaled time.   
The upper curve
depicts evolution of an initial molecular state for $N=40$ and
$|\delta/\Omega|=6$ (below threshold coupling). Clearly the dynamics are
non-periodic. The lower curve shows collapse and revival behaviour for
an initial atomic state for the same parameters.
}
\label{fig3}
\end{figure}

Vardi {\it et al.} \cite{vardi} claim that in the 
large $N$ limit the molecular state is
stationary when $\delta/\Omega=\sqrt{2N}$. Observe however that the molecular
state is never an eigenstate of the Hamiltonian for finite $\delta/\Omega$.
Hence for the molecular state to be stationary it is implied that the
gap between the ground and first excited state closes. 
For a large but finite $N$, the gap does not close (compare Fig. 2).
Our conclusion is that there is   
a smooth transition from the large amplitude, non-periodic coherent
oscillations
(small $\delta/\Omega$)
to the 
quasi-periodic and stable regime (large $\delta/\Omega$), supported by
Figs. 3 and 4.  Our results are in contrast to Fig. 2 in Ref.
\cite{vardi} where the relative population was shown for a sufficiently
short time interval that only the collapse was seen and on this basis it
was mistakenly suggested that this model can describe decoherence
effects.

\begin{figure}[h]
\centerline{
\psfig{file=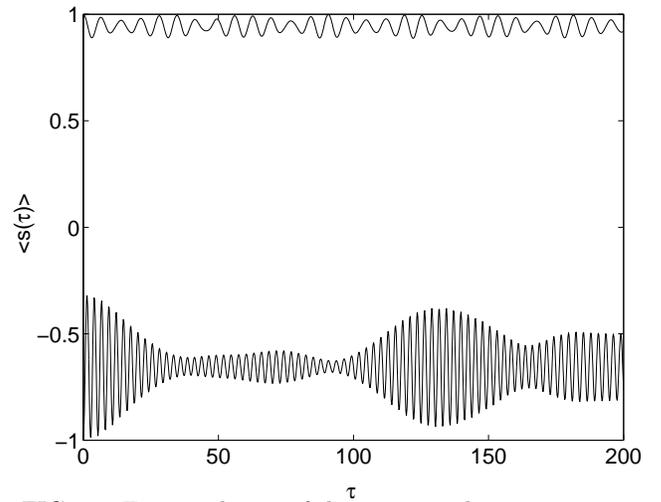,width=8.5cm}}
\caption{
Time evolution of the average relative atom number for $N$=40 and
$|\delta/\Omega|=10$ (above threshold coupling). The upper curve for
an initial molecular state shows stable quasi-periodic behaviour. The lower
curve for an initial atomic state 
illustrates the transition from large amplitude collapse and revival 
to more stable, small amplitude oscillations, as
the detuning $\delta$ is increased.  
}
\label{fig4}
\end{figure}

{\it Conclusion.} 
We have shown that a two-mode model for an atom-molecule BEC
is exactly solvable via the Bethe ansatz. We have conducted both asymptotic
and numerical analysis to establish scaling invariance of the model, 
identified a parity effect in the energy spectrum 
and investigated the quantum dynamics. Our results indicate that the
transition between the atomic and molecular BEC regimes is smooth, in
contrast with Ref. \cite{vardi}. We believe
that the exact solution will help to clarify many facets of the
coherence properties of this model, in order to qualitatively  
compare with current experimental work \cite{wieman}.

We thank Karen Dancer, Peter Drummond and Chris Tisdell  
for valuable discussions. 
This work was supported by the Australian Research Council.
 
\end{multicols}
\end{document}